\documentclass[aps,prb,twocolumn,superscriptaddress]{revtex4-2}
\usepackage{graphicx}
\usepackage{float}
\usepackage{amssymb}
\usepackage{amsmath}
\usepackage{hyperref}
\usepackage{booktabs} 
\usepackage{ulem}      % for \sout
\usepackage{xcolor} % きれいな横線
\usepackage{graphicx}   % 画像を挿入
\usepackage{siunitx}
\usepackage[T1]{fontenc}
\usepackage{lmodern}      % T1 用のフォント（CM の T1 版）。推奨
\usepackage[utf8]{inputenc}

\begin{document}

\title{Raman Spectroscopic Investigation of Ferroaxial Order in Na$_2$BaNi(PO$_4$)$_2$ Single Crystals}

\author{Ryunosuke Takahashi}
\email{rtakahashi@sci.u-hyogo.ac.jp}
\affiliation{Department of Material Science, Graduate School of Science, University of Hyogo, Ako, Hyogo 678-1297, Japan}

\author{Hayato Seno}
\affiliation{Department of Material Science, Graduate School of Science, University of Hyogo, Ako, Hyogo 678-1297, Japan}

\author{Marin Takahashi}
\affiliation{Department of Material Science, Graduate School of Science, University of Hyogo, Ako, Hyogo 678-1297, Japan}

\author{Shigetoshi Tomita}
\affiliation{Department of Material Science, Graduate School of Science, University of Hyogo, Ako, Hyogo 678-1297, Japan}

\author{Reo Fukunaga}
\affiliation{Department of Material Science, Graduate School of Science, University of Hyogo, Ako, Hyogo 678-1297, Japan}

\author{Suguru Nakata}
\affiliation{Department of Material Science, Graduate School of Science, University of Hyogo, Ako, Hyogo 678-1297, Japan}

\author{Takayuki Nagai}
\affiliation{Department of Applied Physics, University of Tokyo, Tokyo 113-8656, Japan}

\author{Shigetada Yamagishi}
\affiliation{Department of Advanced Materials Science, University of Tokyo, Kashiwa, Chiba 277-8561, Japan}

\author{Yoichi Kajita}
\affiliation{Department of Advanced Materials Science, University of Tokyo, Kashiwa, Chiba 277-8561, Japan}

\author{Tsuyoshi Kimura}
\affiliation{Department of Applied Physics, University of Tokyo, Tokyo 113-8656, Japan}

\author{Masami Kanzaki}
\affiliation{Institute for Planetary Materials, Okayama University, Misasa, Tottori 682-0193, Japan}

\author{Hiroki Wadati}
\affiliation{Department of Material Science, Graduate School of Science, University of Hyogo,
Ako, Hyogo 678-1297, Japan}

\begin{abstract}
Ferroaxial order is characterized by the breaking of mirror symmetry parallel to the crystallographic principal axis, which often originates from spontaneous rotational distortions of the crystal lattice. 
Such rotational distortions are, by symmetry, allowed to couple to specific phonon modes. 
However, Raman-active phonons associated with these rotational distortions have not yet been clearly identified on a symmetry-consistent basis. 
Here, we perform polarization-resolved Raman spectroscopy on the ferroaxial phase of Na$_2$BaNi(PO$_4$)$_2$ single crystals and combine the measurements with first-principles lattice-dynamics calculations. 
This symmetry-guided analysis enables a comprehensive assignment of Raman-active 
modes in the ferroaxial phase. Several low-frequency $A_g$ modes exhibit finite 
linewidth broadening, suggesting that these phonons may be weakly affected by the 
underlying rotational distortion. These results establish a symmetry-based spectroscopic 
framework for analyzing phonons associated with rotational distortions in 
ferroaxial materials and provide a basis for future studies of ferroaxial order in complex oxides.

\end{abstract}

\maketitle
\section{Introduction}

Symmetry breaking in crystalline materials is a fundamental mechanism responsible for the emergence of various ordered phases in condensed-matter systems and underlies the development of ferroic functionalities such as ferromagnetism and ferroelectricity.
Ferroaxial order, an axial-vector order often arising from a uniform rotational distortion within a crystal, has recently been recognized as a novel ferroic state~\cite{Johnson2011,Hlinka2016,Yokota2022,HayashidaNatComm2020,HayashidaPRM2021,Jin2020_NatPhys}.
It breaks the mirror symmetry that contains the rotation axis, placing ferroaxial order in a symmetry class distinct from conventional ferroelectric and magnetic orders~\cite{HeKhalsaPRR2024,CaoPRB2014}.
The order parameter is an axial vector—the ferroaxial moment—whose sign reverses between clockwise and counterclockwise rotational distortions, giving rise to domain states.
Glaserite-type compounds with trigonal symmetry~\cite{Nikolova2013} offer model platforms for studying lattice rotations coupled to the ferroaxial moment ~\cite{Yamagishi2023}.
A representative example is RbFe(MoO$_4$)$_2$, which shows symmetry lowering associated with ferroaxial order at $T\mathrm{_C} = 190$ K~\cite{Waskowska2010}.

Recent studies have revealed that ferroaxial moments and domain states can be probed using electrogyration~\cite{HayashidaNatComm2020,HayashidaPRM2021,HayashidaAOM2025}, second-harmonic generation~\cite{Yokota2022}, and circularly polarized Raman scattering~\cite{Liu2023,Kusuno2025}.
Chemical design strategies~\cite{Yamagishi2023,NagaiJACS2023} and observations of unique phenomena such as electric-field-induced magnetochiral dichroism~\cite{HayashidaPNAS2023} and domain switching~\cite{Liu2023,Zeng2025} have further accelerated the field. In 2025, Zeng et al. demonstrated photo-induced switching of ferroaxial order in RbFe(MoO$_4$)$_2$ by generating an effective axial field using circularly driven terahertz phonons~\cite{Zeng2025}. This result shows that ferroaxial degrees of freedom can be optically controlled via phonon excitation in a robust and reversible manner, further highlighting their growing significance as a functional order parameter.

Phonons in ferroaxial phases play a central role in these issues.
Lattice rotations can couple strongly to vibrational modes, producing characteristic signatures in Raman spectra~\cite{KranertPRL2016_AnisotropicRaman,Litasov2017}. However, a complete symmetry-resolved classification of phonons remains lacking~\cite{KranertPRL2016_AnisotropicRaman}. Key issues are
(i) a comprehensive symmetry-consistent assignment of Raman-active phonons in the ferroaxial phase,
(ii) the identification of phonons that are symmetry-allowed to couple to the rotational distortion, and
(iii) the determination of polarization conditions that best isolate such modes.

Na$_2$BaNi(PO$_4$)$_2$ with the glaserite-type structure, the target material of this study, undergoes a continuous symmetry reduction at $T_\mathrm{C} = 746$ K~\cite{KajitaChemMater2024,Yamagishi2023,Nikolova2013}.
Upon cooling, the symmetry lowers from $P\bar{3}m1$ ($D_{3d}$) to $P\bar{3}$ ($S_6$), where mirror symmetry is lost and a ferroaxial moment emerges along the $c$ axis.
This transition reflects collective rotations of PO$_4$ tetrahedra (or equivalently NiO$_6$ octahedra), and right- and left-handed domains appear below $T\mathrm{_C}$~\cite{KajitaJPSJ2025,HayashidaNatComm2020}.

In this study, we performed polarization-resolved Raman spectroscopy on ferroaxial Na$_2$BaNi(PO$_4$)$_2$.
By combining Raman spectra with group-theoretical analysis and first-principles lattice-dynamics calculations, we systematically assigned all Raman-active phonons in the ferroaxial phase.
Among these modes, some low-frequency $A_g$ phonons show relatively broader linewidths. Since our calculations identify an $A_{2g}$ phonon in the high-symmetry phase that evolves into an $A_{g}$ phonon in the ferroaxial phase, this mode becomes symmetry-allowed to couple to the rotational distortion. Therefore, the broadened features observed in this phonon are likely to reflect the influence of the underlying rotational structural degrees of freedom.
These results establish a symmetry-based spectroscopic basis for discussing phonons associated with rotational distortions in ferroaxial materials.

\begin{figure}[t]                   
  \centering

  \includegraphics[width=\columnwidth]{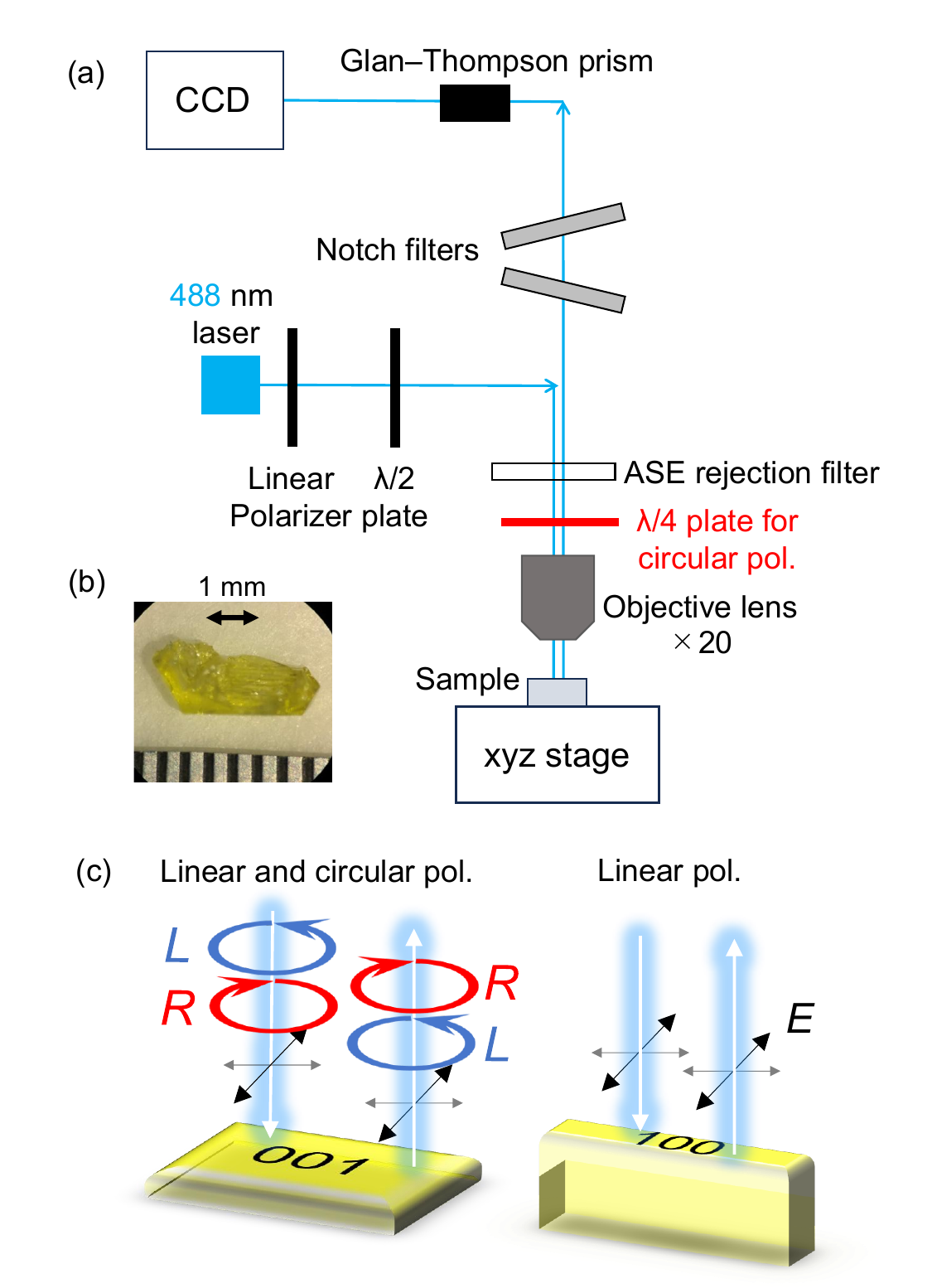}
  \caption{(a) Schematic diagram of the Raman scattering setup. (b) An optical image of a Na\textsubscript{2}BaNi(PO\textsubscript{4})\textsubscript{2} single crystal. (c)Measurement geometry for polarization-resolved Raman spectroscopy, illustrating the incident and scattered light orientations relative to the crystal axes.}
  \label{fig:setup}
\end{figure}

\section{Experimental Methods}

The Raman-scattering intensity is determined by the Raman tensor $\alpha$ imposed by crystal symmetry and by the inner product of $\alpha$ with the Jones vectors, which describe the polarization states of the incident and scattered light.  
For a Raman line to be observed, the following condition must be satisfied:
\[
  I \propto \bigl|\mathbf{e}_\mathrm{s}^{\dagger}\, \alpha\, \mathbf{e}_\mathrm{i}\bigr|^{2},
\]
where $\mathbf{e}_\mathrm{i}$ and $\mathbf{e}_\mathrm{s}$ are the incident and scattered polarization vectors, respectively.

Na$_2$BaNi(PO$_4$)$_2$ belongs to point group $S_{6}$ at room temperature; consequently, the Raman-active phonons exhibit either $A_g$ or $E_g$ symmetry.  
The corresponding Raman tensors are
\begin{subequations}
\begin{align}
\alpha(A_g) &=
\begin{pmatrix}
a & 0 & 0\\
0 & a & 0\\
0 & 0 & b
\end{pmatrix},
\\[6pt]
\alpha(E_g^x) &=
\begin{pmatrix}
c & 0 & 0\\
0 & -c & d\\
0 & d & 0
\end{pmatrix},
\quad
\alpha(E_g^y) &=
\begin{pmatrix}
0 & -c & -d\\
-c & 0 & 0\\
-d & 0 & 0
\end{pmatrix}.
\end{align}
\end{subequations}

The Raman selection rules for the ferroaxial phase ($P\bar{3}$, point group $S_6$) are summarized in Table~\ref{tab:Raman_selection_rules}. In the backscattering geometry $z(\cdot,\cdot)\overline{z}$, the selection rules derived from the Raman tensors indicate that the crossed configuration $z(xy)\overline{z}$ selectively probes the $E_g$ modes, while the parallel configurations, such as $z(xx)\overline{z}$ and $z(yy)\overline{z}$, allow contributions from both $A_g$ and $E_g$. In contrast, in the $x(zz)\overline{x}$ geometry, only the $A_g$ modes are allowed. Similarly, under circular polarization configurations $z(RR)\overline{z}$ and $z(LL)\overline{z}$, only the $A_g$ modes contribute to the Raman intensity. Here, \( R = (\hat{\mathbf{x}} + i\hat{\mathbf{y}})/\sqrt{2} \) and \( L = (\hat{\mathbf{x}} - i\hat{\mathbf{y}})/\sqrt{2} \).  
Co-rotating polarization configurations (\(RR\), \(LL\)) predominantly probe the $A_g$ symmetry modes, whereas counter-rotating ones (\(RL\), \(LR\)) mainly activate the $E_g$ modes, in accordance with the Raman selection rules for the $S_6$ point group.
Thus, the symmetry of each Raman-active mode can be unambiguously identified by comparing the spectra obtained under different polarization geometries.

  \begin{table}[h]
  \centering
  \caption{Raman activity of $A_g$, $E_g^x$, and $E_g^y$ modes for various polarization configurations.}
  \begin{tabular}{lcccc}
    \toprule
    \textbf{Configuration} & \textbf{$A_g$} & \textbf{$E_g^x$} & \textbf{$E_g^y$} & \textbf{Remark} \\
    \midrule
    $z(xx)\bar{z}$   & $\checkmark$ & $\checkmark$ & -- & \textbf{$A_g$} $+$ \textbf{$E_g^x$}\\
    $z(xy)\bar{z}$   & -- & -- & $\checkmark$ &\textbf{$E_g^y$}  \\
    $x(zz)\bar{x}$   & $\checkmark$ & -- & -- &\textbf{$A_g$} \\
    $x(yz)\bar{x}$   & -- & $\checkmark$ & -- & \textbf{$E_g^x$}\\
    $x(zy)\bar{x}$   & -- & $\checkmark$ & -- &\textbf{$E_g^x$} \\
    $x(yy)\bar{x}$   & $\checkmark$ & $\checkmark$ & -- &\textbf{$A_g$} $+$ \textbf{$E_g^x$} \\
    $z(RR)\bar{z}$   & $\checkmark$ & -- & -- &\textbf{$A_g$} \\
    $z(LL)\bar{z}$   & $\checkmark$ & -- & -- &\textbf{$A_g$} \\
    $z(RL)\bar{z}$   & -- & $\checkmark$ & $\checkmark$ &\textbf{$E_g^x$} $+$ \textbf{$E_g^y$} \\
    $z(LR)\bar{z}$   & -- & $\checkmark$ & $\checkmark$ & \textbf{$E_g^x$} $+$ \textbf{$E_g^y$}\\
    \bottomrule
  \end{tabular}
  \label{tab:Raman_selection_rules}
\end{table}

Raman scattering measurements were carried out on Na\textsubscript{2}BaNi(PO\textsubscript{4})\textsubscript{2} single crystals grown by the floating zone method using the experimental setup schematically illustrated in Fig.~\ref{fig:setup} (a).
A semiconductor laser with a wavelength of 488~nm and an output power of 100~mW was employed as the excitation source.
A Coherent optical filter set and a 90/10 beam splitter enabled access to low-frequency signals down to approximately 5~cm$^{-1}$.
The scattered light was passed through two volume holographic notch filters and detected with a spectrometer equipped with a CCD detector.
The polarization-resolved Raman measurement geometry, which indicates the orientations of the incident and scattered light relative to the crystal axes, is illustrated in Fig.~\ref{fig:setup} (b). The four circular polarization configurations ($RL$, $LR$, $RR$, and $LL$) are also defined in the same figure.
An optical image of a Na\textsubscript{2}BaNi(PO\textsubscript{4})\textsubscript{2} single crystal used in the present study is displayed in Fig.~\ref{fig:setup} (c).  

Spectra were recorded on the (001) and (100) facets in the frequency range of 50–1200 cm$^{-1}$.
Both pseudo-depolarized and polarization-resolved configurations were employed.  
In the pseudo-depolarized setup, a quarter-wave ($\lambda/4$) plate (QWP) was inserted to randomize the incident polarization, effectively mimicking a depolarized condition.  
For polarization-resolved measurements, a half-wave ($\lambda/2$) plate (HWP) was additionally used to rotate the incident polarization direction, enabling the selection between parallel (co-polarized) and crossed (orthogonal) geometries according to Porto notation.  
In this notation, $z(xx)\overline{z}$ denotes incident and scattered beams propagating along the $z$ axis, with both polarizations along $x$, whereas $z(xy)\overline{z}$ refers to orthogonal $x$ (incident) and $y$ (scattered) polarizations.  
Circular polarizations were also employed to examine possible chiral contributions associated with ferroaxial domains.

Density functional theory (DFT) calculations were performed on Na$_2$BaNi(PO$_4$)$_2$ using the projector augmented-wave method \cite{Blochl1994,Kresse1999} as implemented in the VASP code \cite{Kresse1996a,Kresse1996b}. All calculations employed the Perdew–Burke–Ernzerhof form of the generalized gradient approximation optimized for solids (GGA–PBEsol) to treat exchange–correlation interactions \cite{Perdew2008}. The following states were treated as valence electrons: 2$p$ and 3$s$ for Na; 5$s$, 5$p$, and 6$s$ for Ba; 3$d$ and 4$s$ for Ni; 3$s$ and 3$p$ for P; and 2$s$ and 2$p$ for O. The on-site Hubbard $U$ correction \cite{Dudarev1998} was applied to the Ni 3$d$ orbitals with an effective $U$ value of 6.0 eV \cite{Zhou2004}. A plane wave cutoff energy of 550 eV was used, and all plane waves with kinetic energies below this value were included in the basis set. The $\Gamma$-centered mesh sampling in the Brillouin zone was set to $7 \times 7 \times 5$ for structural optimization and $15 \times 15 \times 12$ for the density of states calculation. The structural optimization was performed until the residual stress and forces were less than 0.01 GPa and 1 meV/Å, respectively. 
Phonon band structures were calculated for $2 \times 2 \times 2$ supercells of the $P\bar{3}$ and $P\bar{3}m1$ structures, and the force constants were obtained using the PHONOPY package \cite{Togo2008}.

Electronic states were analyzed using X-ray photoemission spectroscopy (XPS)  with a ULVAC-PHI PHI 5000 VersaProbe III system equipped with a monochromatized Al K$\alpha$ source (h$\nu$ = 1486.8 eV).
The overall energy resolution, estimated from the Ag Fermi edge reported in Ref.~\cite{TakahashiAPL2021}, was approximately 530~meV. To prevent surface charging during XPS measurements, a low-energy electron gun was employed for charge neutralization.  
The binding energies were calibrated with reference to the C~1$s$ peak at 284.8~eV, originating from adventitious carbon \cite{Moulder1992}.

\section*{Results and Discussion}
%======================================================================
%  X-ray Photoelectron Spectroscopy (XPS)
%======================================================================

Before discussing the phonon behavior in detail, we verified the sample composition and chemical states by XPS to ensure the intrinsic stoichiometry of the studied crystals.  
We then established the polarization selection rules, assigned the Raman-active phonons, and benchmarked the observed frequencies against first-principles calculations.
 
\begin{figure}[h]
  \centering
  \includegraphics[width=\columnwidth]{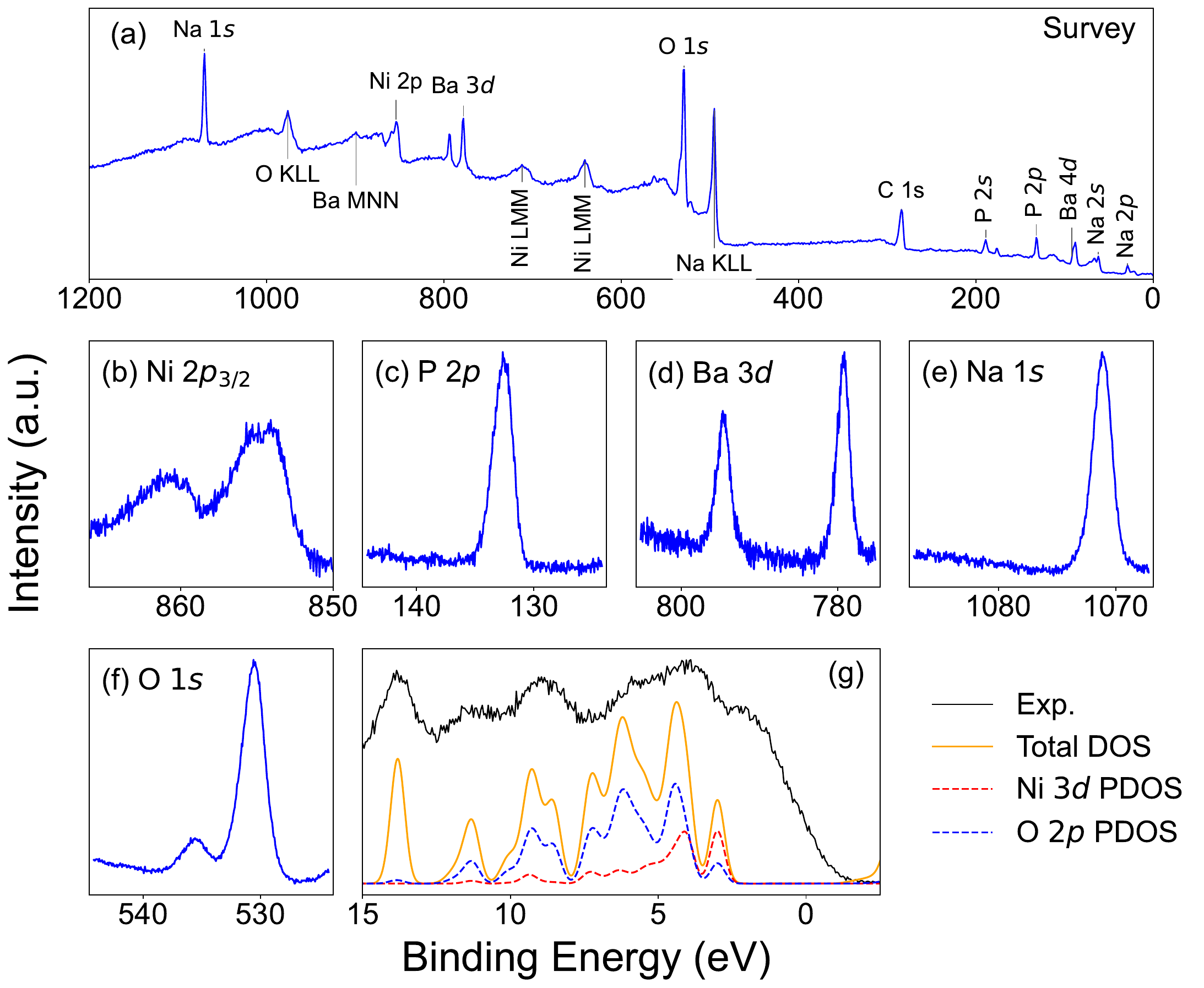}
  \caption{Core-level and valence-band XPS spectra of Na$_2$BaNi(PO$_4$)$_2$ single crystal.
  Panels~(a)--(f) show core-level XPS spectra, while panel~(g) compares the valence-band XPS spectrum (black)
  with the calculated partial density of states (PDOS, colored curves).
  The PDOS was broadened using a Gaussian function with a width of 530~meV to simulate
  the experimental energy resolution.}
  \label{fig:XPS}
\end{figure}

Figure~\ref{fig:XPS} (a-f) shows the survey and core-level spectra.  
The wide-energy survey spectrum contains Na~1$s$, Ni~2$p$, Ba~3$d$, O~1$s$, and P~2$p$ signals, along with the C~1$s$ peak, confirming that all target elements are present. Narrowly scanned spectra were collected for each core level.  
The Ni~2$p$ spectrum exhibits a well-defined 2$p$$_{3/2}$ peak at 854.8~eV and a corresponding 2$p$$_{1/2}$ component with a pronounced satellite at 860.8~eV, consistent with the Ni$^{2+}$ oxidation state.  
The O~1$s$ spectrum shows a dominant peak at 530.65~eV, attributed to lattice O$^{2-}$, and a weak higher-energy shoulder at 535.5~eV arising from surface contamination or adsorbed species.  
The P~2$p$ line appears at 132.65~eV, characteristic of P$^{5+}$ in phosphate groups \cite{Moulder1992}.  
Clear Na~1$s$ and Ba~3$d$ signals are also observed at 1071.1~eV and 779.3/794.6~eV (3$d$$_{5/2}$/3$d$$_{3/2}$), respectively, consistent with Na$^{+}$ and Ba$^{2+}$ cations \cite{Moulder1992}. These XPS results confirm that the Na$_2$BaNi(PO$_4$)$_2$ crystals contain all the designed elements in their expected oxidation states: Na$^{+}$, Ba$^{2+}$, Ni$^{2+}$, P$^{5+}$, and O$^{2-}$.

\begin{figure*}[ht]
  \centering
  \includegraphics[width=\linewidth]{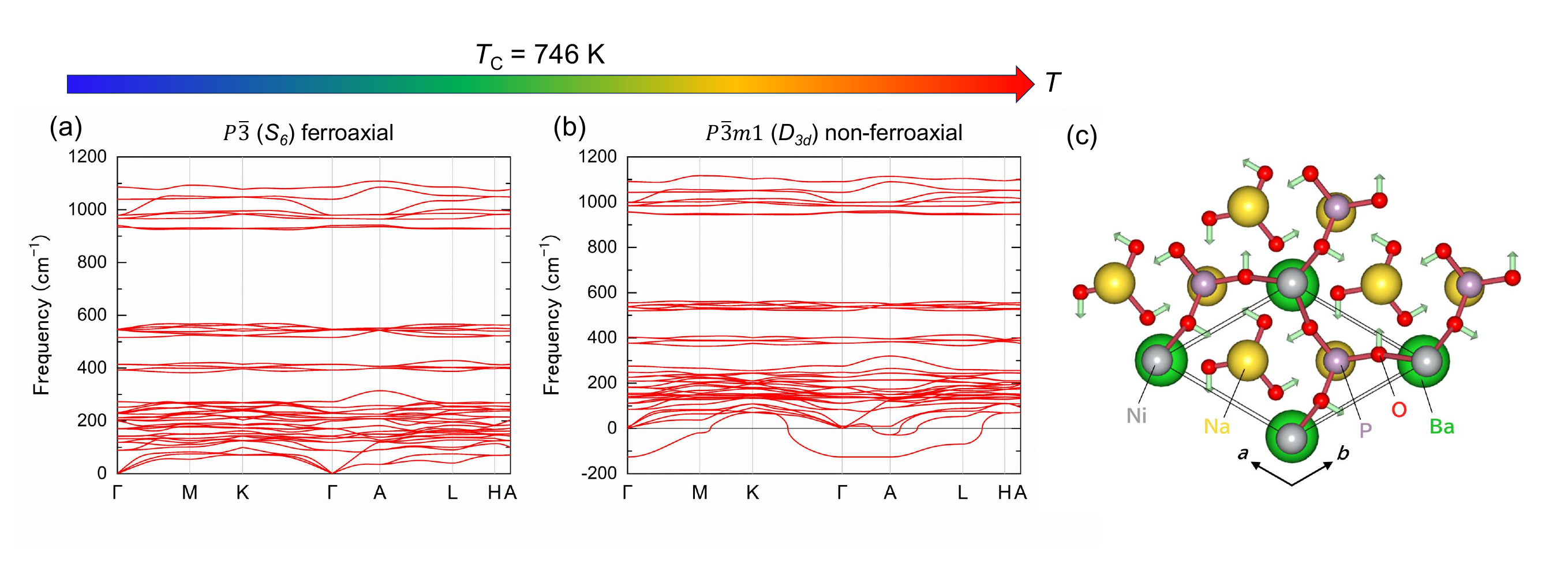}
  \caption{
    First-principles phonon dispersion of Na$_2$BaNi(PO$_4$)$_2$:  
    (a) low-temperature ferroaxial phase (space group $P\overline{3}$, point group $S_6$);  
    (b) high-temperature non-ferroaxial phase (space group $P\overline{3}m1$, point group $D_{3d}$);  
    and (c) atomic displacement pattern of the zone-center $A_{2g}$ phonon obtained from DFT calculations.  
  }
  \label{fig:phonon_dispersion}
\end{figure*}
Figure~\ref{fig:XPS} (g) compares the valence-band XPS spectrum of Na$_2$BaNi(PO$_4$)$_2$ with the density of states (DOS) obtained from first-principles calculations.  
The binding-energy scale of the experimental spectrum was calibrated with reference to the C~1$s$ peak at 284.8~eV.  
In the theoretical partial density of states (PDOS), a Gaussian broadening of 530 meV was applied to account for the experimental resolution.  
The calculated PDOS spans the range from $-2$ to 15~eV and is dominated by Ni~3$d$ states.  
The experimental spectrum shows good agreement with the calculations in both the peak positions and relative intensities, particularly near the Fermi level. The finite intensity observed near 0 eV in the experimental valence-band spectrum does not originate from metallic behavior. Instead, it is caused by spectral broadening due to charge neutralization with the electron flood gun, as the sample is insulating. 
These results confirm that the electronic structure of the sample is consistent with stoichiometric Na$_2$BaNi(PO$_4$)$_2$, validating the use of DFT calculations for the subsequent phonon analysis.

To clarify the lattice-dynamical origin of the ferroaxial phase transition and to compare it with the experimental Raman spectra, phonon dispersion calculations were performed for both the low-temperature ferroaxial phase (space group $P\bar{3}$, point group $S_6$) and the high-temperature non-ferroaxial phase (space group $P\bar{3}m1$, point group $D_{3d}$). The calculated phonon dispersion curves are shown in Fig. \ref{fig:phonon_dispersion} (a, b), where imaginary frequencies are plotted along the negative axes. In the non-ferroaxial phase, an unstable optical phonon mode with a significant imaginary frequency appears at the $\Gamma$ point, which transforms as the irreducible representation (Irrep) $A_{2g}$, indicating a vibrational instability in this phase. The atomic displacement pattern of this mode is shown in Fig. \ref{fig:phonon_dispersion} (c). The eigenvector of this $A_{2g}$ unstable mode involves ferri-like rotational displacements, comprising the rotation of NiO$_6$ octahedra coupled with counter-rotations of the PO$_4$ tetrahedra, which are consistent with the atomic displacement pattern observed across the ferroaxial phase transition. These results indicate that an instability-driving phonon mode plays a key role in stabilizing the ferroaxial structure of Na$_2$BaNi(PO$_4$)$_2$.
By analogy to displacive ferroelectric transitions, the emergence of soft-mode-like dynamics may be anticipated. Although additional unstable modes are also found at the zone boundary (e.g., at the $A$ and $L$ points), freezing these modes would result in superlattice structures that have not been observed experimentally. In contrast, in the ferroaxial phase, the soft modes disappear, confirming the ferroaxial $P\bar{3}$ structure as the crystallographic ground state for Na$_2$BaNi(PO$_4$)$_2$.
The symmetry reduction from $P\bar{3}m1$ to $P\bar{3}$ allows an $A_{2g}$ phonon in the high-symmetry phase to evolve into an $A_{g}$ phonon in the ferroaxial phase, indicating that this vibrational mode is symmetry-allowed to couple to the rotational distortion.
Therefore, it is important to investigate A$_g$ modes in the ferroaxial phase in Na$_2$BaNi(PO$_4$)$_2$. 

\begin{table}[h]
  \caption{Calculated and experimental Raman-active phonons of Na$_2$BaNi(PO$_4$)$_2$ in the ferroaxial ($S_6$) phase.}
  \label{tab:modes}
  \centering
  \begin{tabular}{@{}c S[table-format=4.0] c@{}}
    \toprule\toprule
    Irrep & {Calculated (cm$^{-1}$)} & {Experimental (cm$^{-1}$)} \\
    \midrule
    $E_g$ &  119 & 117 \\ 
    $A_{g}$ &  133 & 140 \\
    $A_{g}$ &  142 & 154 \\
    $E_g$ &  205 & 207 \\
    $E_g$ &  229 & 225 \\
    $A_{g}$ &  252 & 250 \\
    $E_g$ &  414 & 434 \\
    $A_{g}$ &  545 & 578 \\
    $E_g$ &  548 & 578 \\
    $A_{g}$ &  940 & 977 \\
    $E_g$ &  967 & 994 \\
    $A_{g}$ & 1086 & 1117 \\
    \bottomrule\bottomrule
  \end{tabular}
\end{table}

\begin{figure*}[h]
  \centering
  \includegraphics[width=\linewidth]{"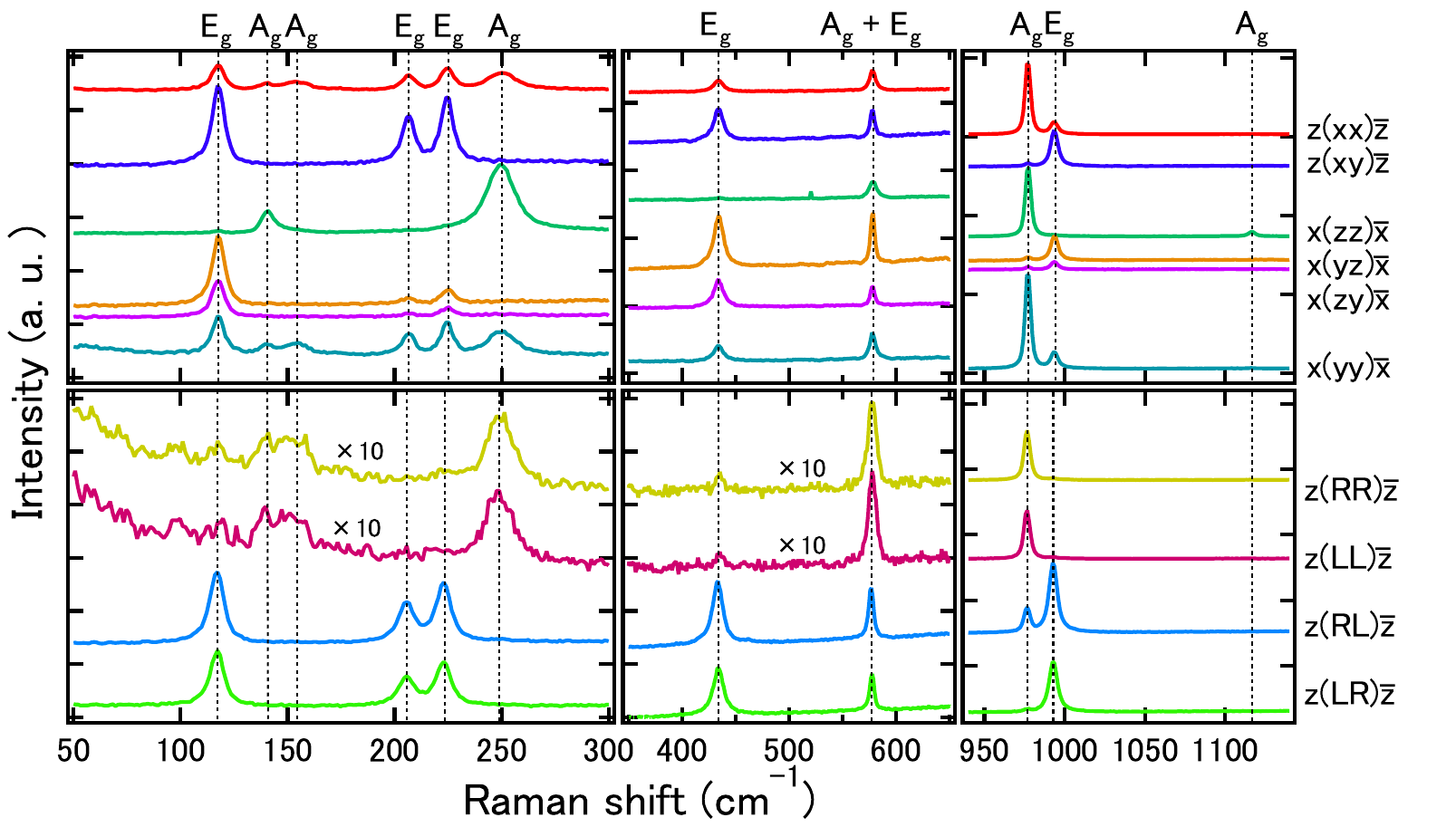"}
  \caption{
Polarization-dependent Raman spectra of Na$_2$BaNi(PO$_4$)$_2$ single crystals at room temperature.  
Raman spectra were measured under ten Porto-notation polarization configurations.  
The figure displays the characteristic spectral regions containing distinct Raman peaks,  
covering the ranges of 50–300, 350–650, and 940–1140~cm$^{-1}$.  
For clarity, some spectra are vertically offset, and the $RR$ and $LL$ spectra are magnified by a factor of ten due to their relatively weak intensities.
  }
  \label{fig:raman_temp}
\end{figure*}

Figure~\ref{fig:raman_temp} shows the polarization-dependent Raman spectra of Na$_2$BaNi(PO$_4$)$_2$ single crystals.
Measurements were performed under ten Porto-notation configurations, and Raman features were observed in the ranges of 50–300, 350–650, and 940–1140~cm$^{-1}$.
The $RR$ and $LL$ spectra are magnified tenfold for clarity.

To examine the mode assignments, the Raman peak positions obtained under the $z(xx)\overline{z}$ configuration—where both $A_g$ and $E_g$ modes are observable—were compared with the phonon frequencies calculated at the $\Gamma$ point. We fitted the experimental data with a Lorentzian function to estimate the peak position, as summarized in Appendix~A (Table \ref{tab:A1}).
The Raman-active modes summarized in Table~\ref{tab:modes} were used as a reference for assigning the observed peaks.
The calculated and experimental frequencies show a broad correspondence, which is consistent with the symmetry assignments expected for the ferroaxial ($S_6$) phase. Thus, the polarization dependence further distinguishes the $A_g$ and $E_g$ symmetry components, providing additional support for the mode assignments based on the expected selection rules. The $A_g$ mode at 977 cm$^{-1}$ appears in the $z(RL)\bar{z}$ configuration, most likely because this peak is intrinsically strong and the circular polarization is not perfectly pure. Elliptically polarized light inevitably contains a linear polarization component, which can partially activate the $A_g$ mode through the corresponding selection rules. In addition, even under linear‐polarization configurations, slight misalignment between the crystal axes and the polarization direction can satisfy the selection rules, causing modes that are not expected in this geometry to appear with very weak intensity.

\begin{figure}
    \centering
    \includegraphics[width=\linewidth]{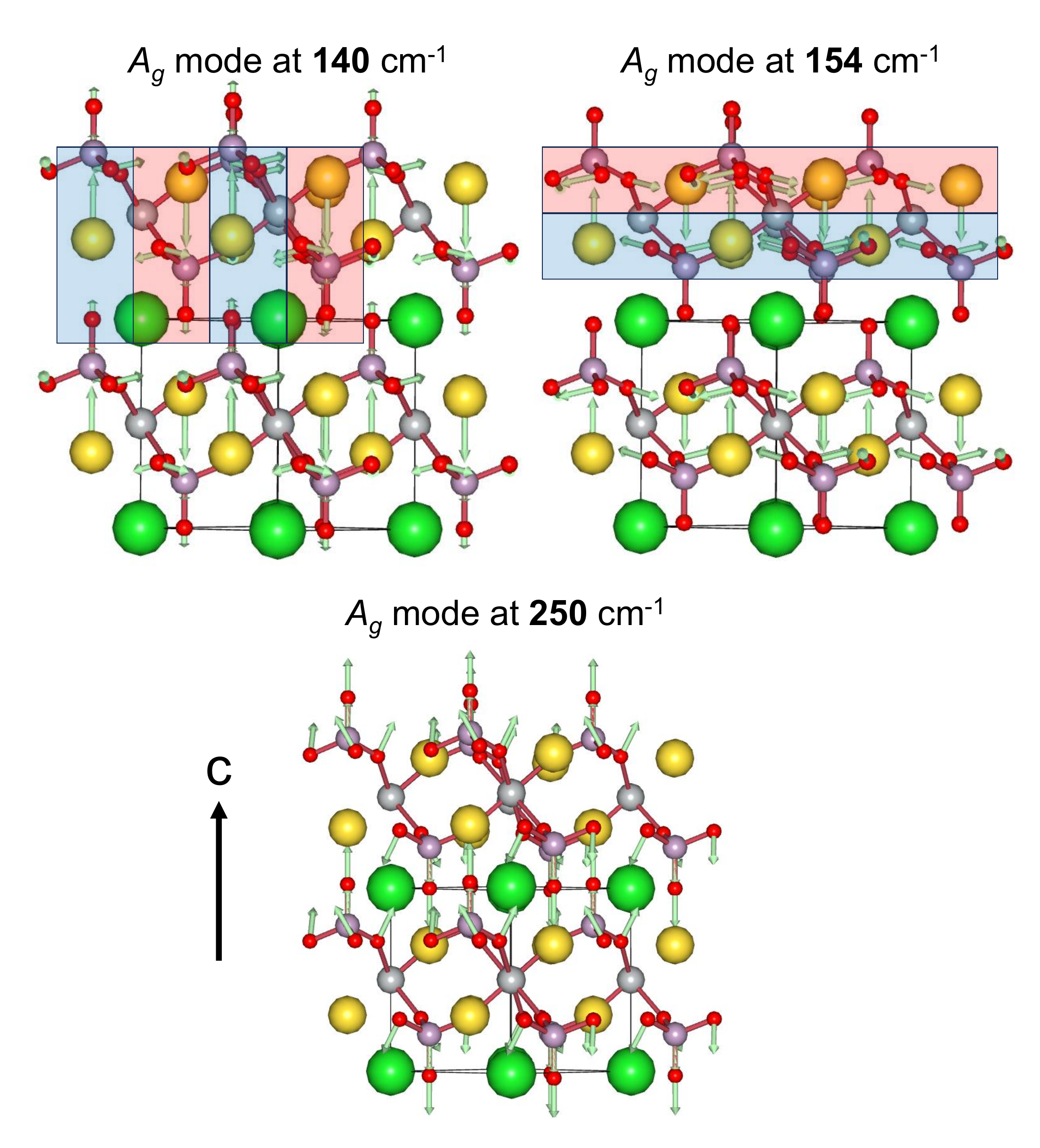}
    \caption{Calculated eigenvectors of the low-frequency $A_g$ phonon modes in Na$_2$BaNi(PO$_4$)$_2$ at 140, 154, and 250~cm$^{-1}$.
    Arrows indicate atomic displacement directions.
    Red and blue shaded regions denote parts of the structure where the atomic displacements have the same or opposite sign of the $c$-axis component, respectively, visualizing the degree of cancellation or collectivity of the out-of-plane motion.}
    \label{fig:Ag_mode}
\end{figure}

The linewidths of Raman peaks were estimated using Lorentzian fitting, and the resulting parameters are summarized in Table~\ref{tab:A1}.
Among the $A_g$ modes below 300~cm$^{-1}$, the modes at 154 and 250~cm$^{-1}$ exhibit relatively large linewidths of 6.8 and 7.8~cm$^{-1}$, respectively, whereas the $E_g$ modes in the same frequency range tend to show narrower linewidths.
In contrast, the $A_g$ mode at 140~cm$^{-1}$ is fitted with a comparatively narrow linewidth, comparable to those of the $E_g$ modes.
The observed variation in linewidths indicates a clear mode dependence in the low-frequency region, although in-plane rotational displacements are present in all three modes at 140, 154, and 250 cm$^{-1}$.
While Raman linewidths can be influenced by spectral overlap, instrumental resolution, and fitting uncertainties, the systematic difference between the $A_g$ and $E_g$ modes suggests that intrinsic lattice-dynamical factors may also play a role.
This mode selectivity is consistent with the distinct eigenvector characteristics obtained from lattice-dynamical calculations.
Specifically, the 140~cm$^{-1}$ mode involves locally canceling out-of-plane displacements, characterized by largely antiparallel motions of Na and oxygen ions.
By contrast, the $A_g$ modes at 154 and 250~cm$^{-1}$ exhibit more collective displacements associated with rotational distortions of the lattice, as illustrated in Fig.~\ref{fig:Ag_mode}.
The 250~cm$^{-1}$ mode further involves internal deformations of the polyhedral units.
Such internal distortions are generally expected to increase the sensitivity of optical phonons to anharmonic decay processes, which are known to influence Raman linewidths \cite{Klemens1966}.
In Na$_2$BaNi(PO$_4$)$_2$, low-energy structural degrees of freedom related to rotational distortions are suggested by symmetry considerations and lattice-dynamical calculations, including the predicted $A_{2g}$ mode in the high-temperature phase (Fig.~\ref{fig:phonon_dispersion}(b)).
Although a quantitative assessment of the dynamical contributions of these modes is beyond the scope of the present work, the presence of such low-energy rotational degrees of freedom provides a plausible context for discussing the mode-dependent linewidth variation observed among the low-frequency $A_g$ phonons.

In contrast, at higher frequencies where the internal vibrations of the PO$_4$ tetrahedra dominate, certain $A_g$ modes remain comparatively sharp, reversing the linewidth trend observed in the low-frequency region.
Above 950~cm$^{-1}$, the Raman features originate from the internal stretching vibrations of the (PO$_4)^{3-}$ tetrahedra.
The sharp peak at 977~cm$^{-1}$ is assigned to the $\nu_1$ ($A_g$) symmetric stretching mode, whereas the broader band at 994~cm$^{-1}$ corresponds to the $\nu_3$ ($E_g$) antisymmetric stretching mode.
These assignments are consistent with previous studies on phosphate compounds, where the $\nu_1$ and $\nu_3$ modes typically appear in the 950–990 and 1000–1150~cm$^{-1}$ ranges, respectively~\cite{Jirasek2017,Litasov2017}.
Lorentzian fitting yields half widths at half maximum of 3.81~cm$^{-1}$ for the $\nu_1$ ($A_g$) mode and 5.3~cm$^{-1}$ for the $\nu_3$ ($E_g$) mode (Table~\ref{tab:A1}).
The comparatively narrower linewidth of the $\nu_1$ mode is consistent with the totally symmetric stretching vibration being less influenced by crystal-field perturbations and local structural disorder than the antisymmetric $\nu_3$ vibration~\cite{Nakamoto2009}.

\section*{Conclusion}

In this work, we established a symmetry-guided and comprehensive framework for identifying Raman-active phonons in Na$_2$BaNi(PO$_4$)$_2$ single crystals. By combining polarization-resolved Raman spectroscopy with group-theoretical tensor analysis and first-principles lattice-dynamics calculations, we achieved a systematic symmetry assignment of all Raman-active modes. Although the ferroaxial-related $A_g$ mode at room temperature exhibits spectral overlap with broadened $A_g$ phonons due to finite phonon lifetimes, along with residual $E_g$ components originating from the finite extinction ratio of the optics, slight crystal misalignment, and polarization mixing intrinsic to micro-Raman measurements with a finite numerical aperture, the polarization-resolved Raman analysis nevertheless enabled a reliable and detailed mode assignment.
Because $T_\mathrm{C}$ exceeds the temperature range accessible to conventional Raman setups,
direct tracking of phonon modes that may couple to the rotational distortion is challenging in the present experiment,
highlighting the need for future high-temperature Raman measurements.

These findings establish a symmetry-grounded spectroscopic framework for identifying and analyzing phonons linked to mirror distortions in ferroaxial materials and lay the groundwork for uncovering how rotational symmetry breaking ascribed to rotational distortions shapes the lattice dynamics of complex oxides.

\section*{Acknowledgments}
This work was supported by the MEXT Quantum Leap Flagship Program (MEXT Q-LEAP) Grant No. JPMXS0118068681 and JSPS KAKENHI Grant-in-Aid Nos. JP23K2580, JP24K08561, JP25H01251, JP25H01247, and JP25H00392.
Low-frequency Raman measurements were performed using the joint-use facilities of the Institute for Planetary Materials at Okayama University.
S.N. is grateful to the Murata Science Foundation, the Hyogo Science and Technology Association, and the Nippon Sheet Glass Foundation for Materials Science and Engineering for their financial support.

\appendix
\setcounter{figure}{0}
\setcounter{table}{0}
\setcounter{equation}{0}
\renewcommand{\thefigure}{A\arabic{figure}}
\renewcommand{\thetable}{A\arabic{table}}
\renewcommand{\theequation}{A\arabic{equation}}
\section*{Appendix A: Lorentzian decomposition and analysis of Raman spectra}

The Raman spectra of Na$_2$BaNi(PO$_4$)$_2$ single crystals were analyzed 
by fitting each peak with a sum of Lorentzian functions to quantitatively 
determine the peak positions and linewidths. 
The fitting function is expressed as
\begin{equation}
I(x) = C + \sum_{i=1}^{N} \frac{A_i \gamma_i^2}{(x - x_{0,i})^2 + \gamma_i^2},
\end{equation}
where $C$ is a constant offset, 
$A_i$ is the amplitude of the $i$th component, 
$x_{0,i}$ is the peak position, 
and $\gamma_i$ is the half width at half maximum.
The optimal parameters were obtained using the least-squares method 
for several frequency regions covering low, mid, and high Raman shifts.

\begin{figure*}[h]
  \centering
  \includegraphics[width=\linewidth]{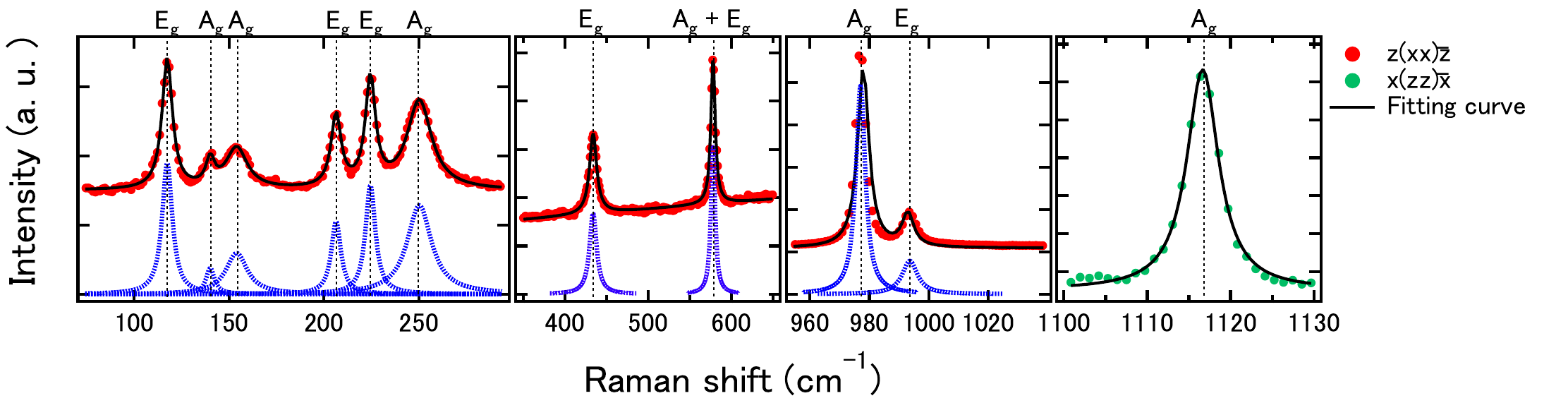}
  \caption{
    Raman spectra of Na$_2$BaNi(PO$_4$)$_2$ single crystals.
    The figure consists of three panels showing the low-, mid-, and high-frequency ranges measured in the $z(xx)\bar{z}$ configuration, and an additional panel showing the high-frequency range measured in the $x(zz)\bar{x}$ configuration.
    The red dots represent the experimental data obtained in the $z(xx)\bar{z}$ geometry, while the green dots correspond to the data measured in the $x(zz)\bar{x}$ geometry.
    The black lines show the total fitted curves, and the blue dashed lines represent the individual Lorentzian components.
    The symmetry assignments ($E_g$, $A_g$) and the vertical dashed lines indicate the mode symmetries and the peak positions, respectively.
  }
  \label{fig:A1}
\end{figure*}

Figure~\ref{fig:A1} shows representative fitting results for the low
(50–300~cm$^{-1}$), mid (400–600~cm$^{-1}$), and high-frequency
(960–1020~cm$^{-1}$) ranges.
All spectra were fitted using the data obtained in the
$z(xx)\bar{z}$ scattering configuration.
For the $A_g$ mode located at 1117~cm$^{-1}$, the fitting was
performed using the $x(zz)\bar{x}$ spectrum, since this mode exhibits a
significantly stronger intensity in that geometry.
The fitting in this case was carried out over the 1100–1130~cm$^{-1}$
range.
The red dots indicate experimental data, the black line denotes the total fit, 
and the blue dashed lines represent individual Lorentzian components. 
The labels ($E_g$, $A_g$) and vertical dashed lines indicate 
the symmetry assignment and peak positions, respectively.

The fitting parameters obtained for all frequency ranges 
are summarized in Table~A1. 
A six-component Lorentzian model reproduces the complex line shape 
in the low-frequency region, while two-component models are sufficient 
for the mid- and high-frequency regions.
Importantly, in the low-frequency range, 
the linewidths $\gamma_i$ of the $E_g$ modes are systematically broader than those of the $A_g$ modes.
\begin{table}[h]
  \caption{Lorentzian fitting parameters of Raman-active modes in Na$_2$BaNi(PO$_4$)$_2$. 
  Uncertainties in parentheses represent one standard deviation.}
  \label{tab:A1}
  \centering
  \begin{tabular}{@{}cccc@{}}
    \toprule
    Component & $x_{0,i}$ (cm$^{-1}$) & $\gamma_i$ (cm$^{-1}$) & Symmetry \\
    \midrule
    % --- Low-frequency region (this work)
    1 & 117  & 3.27(8)  & $E_g$ \\
    2 & 140 & 2.73(41) & $A_g$ \\
    3 & 154 & 6.83(45) & $A_g$ \\
    4 & 206  & 3.14(14) & $E_g$ \\
    5 & 225  & 3.17(10) & $E_g$ \\
    6 & 250 & 7.81(19) & $A_g$ \\
    \midrule
    % --- Mid-frequency region (this work)
    7 & 434        & 10.36(23) & $E_g$ \\
    8 & 578        & 6.074(91) & $A_g$+$E_g$ \\
    \midrule
    % --- High-frequency region (this work)
    9  & 977 & 3.81(5)  & $A_g$ \\
    10 & 993  & 5.30(37) & $E_g$ \\
    11 & 1117 & 4.70(11) & $A_g$ \\
    \bottomrule
  \end{tabular}
\end{table}

\bibliography{references}

\end{document}